\documentclass[%
aps,twocolumn,superscriptaddress,
 amsmath,amssymb,
]{revtex4-2}

\usepackage{graphicx}
\usepackage{dcolumn}
\usepackage{bm}
\usepackage{rotating}
\usepackage[section]{placeins}
\usepackage{tabularx}
\usepackage{enumitem}

\usepackage[utf8]{inputenc}
\usepackage[T1]{fontenc}
\usepackage{mathptmx}
\usepackage{caption}
\usepackage{subcaption}
\begin{document}

\title[S. Mumford et al.]{Emergence of Ferromagnetism Through the Metal-Insulator Transition in Undoped Indium Tin Oxide Films}

\author{Samuel Mumford}
\affiliation{Geballe Laboratory for Advanced Materials, Stanford University, Stanford CA, 94305, USA}
\affiliation{Department of Physics, Stanford University, Stanford CA, 94305, USA}
\author{Tiffany Paul}%
\affiliation{Geballe Laboratory for Advanced Materials, Stanford University, Stanford CA, 94305, USA}
\affiliation{Department of Applied Physics, Stanford University, Stanford CA, 94305, USA}
\author{Aharon Kapitulnik}%
\affiliation{Geballe Laboratory for Advanced Materials, Stanford University, Stanford CA, 94305, USA}
\affiliation{Department of Physics, Stanford University, Stanford CA, 94305, USA}
\affiliation{Department of Applied Physics, Stanford University, Stanford CA, 94305, USA}%

\date{\today}

\begin{abstract}
We present a detailed study of the emergence of bulk ferromagnetism in low carrier density samples of undoped indium tin oxide (ITO). We used annealing to increase the density of oxygen vacancies and change sample morphology without introducing impurities through the metal insulator transition (MIT). We utilized a novel and highly sensitive ``Corbino-disk torque magnetometry'' technique to simultaneously measure the thermodynamic and transport effects of magnetism on the same sample after successive annealing. With increased sample granularity, carrier density increased, the sample became more metallic, and ferromagnetism appeared as resistance approached the MIT. Ferromagnetism was observed through the detection of magnetization hysteresis, anomalous Hall effect (AHE), and hysteretic magnetoresistance. A sign change of the AHE as the MIT is approached may elucidate the interplay between the impurity band and the conduction band in the weakly insulating side of the MIT. \footnote{Corresponding author is Samuel Mumford at smumfor2@stanford.edu}
\end{abstract}
\maketitle

\section{Introduction}

The initial discovery of room-temperature ferromagnetism in Co-doped TiO$_2$ \cite{matsumoto_2001} films was soon followed by the observation of similar behavior in other transition-metal (TM) doped oxides including Co-doped SnO$_2$ \cite{SnO2}, Mn-doped ZnO \cite{ZNO_2003}, Co-doped CeO$_2$ \cite{CeO2_2006}, and Ni, Mo, Fe, and Mn-doped In$_2$O$_3$ \cite{hong_2005,peleckis_2006,Fe_In2O3,Mo_In2O3}. The origin of the observed thin film magnetism continues to be debated, particularly after the discovery of ferromagnetism in undoped oxides such as HfO$_2$ \cite{HfO2}, TiO$_2$, and In$_2$O$_3$ \cite{Hong_2006}. At the heart of the puzzle is observation that the respective bulk undoped oxides exhibit only diamagnetism. Similarly, doping level in TM doped oxides was often too low to explain the observed strength, anisotropy, or thermal treatment sensitivity of the magnetic state. Correlation between the samples' morphology and the occurrence of magnetism instead suggests an explanation based on a spontaneous formation of defects or oxygen vacancies \cite{Hong_2006}. The abundant surface vacancies of the nanoparticles which make up a granular structure provide unpaired electron spins which could interact via an exchange mechanism \cite{main_fm}.

With the observation of ferromagnetism in In$_2$O$_3$ and SnO$_2$, it is natural to explore whether the widely used solid-solution of tin-doped indium oxide (ITO) exhibits ferromagnetism. Ferromagnetism could be used to integrate this optoelectronic material into novel spintronics applications. Indeed, various TM doped \cite{philip_2004, peleckis_2006, ohno_2006, venkatesan_2008, kim_ji_2006} and undoped \cite{hakimi_paper, hong_2005, majumdar_ITO} ITO systems were shown to exhibit ferromagnetism that persisted to room temperature. Oxygen vacancy defects have been proposed as the source of magnetic moments rather than impurities \cite{coey_2010, majumdar_ITO,hakimi_ch5}, while itinerant carriers mediate the collective ferromagnetic state \cite{kim_ji_2006,Stankiewicz_2007,hakimi_paper}. A persistent obstacle in proclaiming intrinsic ferromagnetism for ITO has been the difficulty in observing the anomalous Hall effect (AHE) and hysteretic magnetoresistance. For example, in Cr-doped ITO AHE appears only at very high carrier concentrations exceeding 10$^{21}$~cm$^{-3}$ \cite{kim_ji_2006}, while in 12\% Co-doped ITO (10\% Sn) AHE appears above $\sim 10^{19}$~cm$^{-3}$. Samples of ITO with lower concentrations of magnetic elements generally do not show AHE, and the observed negative magnetoresistance is attributed to magnetic-field induced reduction in spin scattering \cite{kim_ji_2006,hakimi_ch5,Stankiewicz_2007}. Similar to other magnetic semiconductors (e.g. \cite{Arsenault_2008}), ITO is a heavily doped $n$-type semiconductor typically studied near the metal-insulator transition (MIT).  In particular,  undoped ITO close to the MIT may exhibit local magnetic moments that are typically self-generated in quenched disordered electronic systems due to interaction effects \cite{Paalanen1986}. Thus carrier density may not be the only relevant parameter for magnetism \cite{Burkov_2003,Stankiewicz_2007}. It is correspondingly important to elucidate the occurrence of ferromagnetism in undoped ITO within these contexts.

In this work we examine the interplay between morphology, proximity to the MIT, and the emergence of bulk magnetic properties in low carrier density samples of ITO. We used annealing to increase the density of oxygen vacancies and change sample granularity without introducing impurities, and annealed through the MIT \cite{SEM_paper, hakimi_ch4, RTA_ITO, Lee_grain,kerkache2007,AFM,gulen_2012, wu_chiou_1993}. We utilized a novel and highly sensitive ``Corbino-disk torque magnetometry'' technique to simultaneously measure the thermodynamic and transport effects of magnetism on the same sample after successive annealing. Our results clearly show that: \textit{i}) Starting with a highly insulating mostly amorphous ITO with no observed magnetism, successive annealing induces morphology change towards more granular structure. This change is accompanied by increase in carrier density and occurrence of ferromagnetism near the MIT. \textit{ii}) The amplitude of magnetization hysteresis and saturating field both increase with decreased in sheet-resistance, indicating strengthening of the ferromagnetic state. \textit{iii}) Undoped ITO thus is ferromagnetic, exhibiting anomalous Hall effect and hysteretic magnetoresistance for carrier density as low as $\approx 3\times 10^{18}$ carriers/cm$^3$. \textit{iv}) Both AHE and hysteretic magnetoresistance are observed across the MIT. The sign of AHE changes from negative on the insulating side of the MIT to positive AHE on the metallic side of the MIT.

\section{Experiment}

\subsection{Preparation of Samples}

Hall bar and Corbino disk patterned ITO samples were prepared by RF magnetron sputtering using a 10$\%$ Sn target at 8~W/Inch$^2$, 5~mTorr of Ar pressure, and 2.5$\%$ partial pressure of O$_{2}$. Both Hall bar and Corbino disk samples were subjected to successive annealing schedules at 400~K and pressure of $<~10^{-5}~$Torr with measurements performed after each anneal. A typical ITO sample was initially highly insulating and exhibited a starting sheet resistance of 70~k$\Omega/\square$ at room temperature and a thickness of 40~nm. Resistance increased to 460~k$\Omega/\square$ at the measurement base temperature of 4.2~K. Measurements were performed in a liquid helium cryostat with a $7~$T superconducting magnet at temperature close to base temperature. Hall effect and magnetoresistance were measured on the Hall bars using standard 5 lead configuration. The Corbino disk samples were measured by torque magnetometry as described below.

\subsection{Corbino Disk Torque Magnetometry}

\begin{figure}[h]
     \centering
         \includegraphics[width=.45\textwidth]{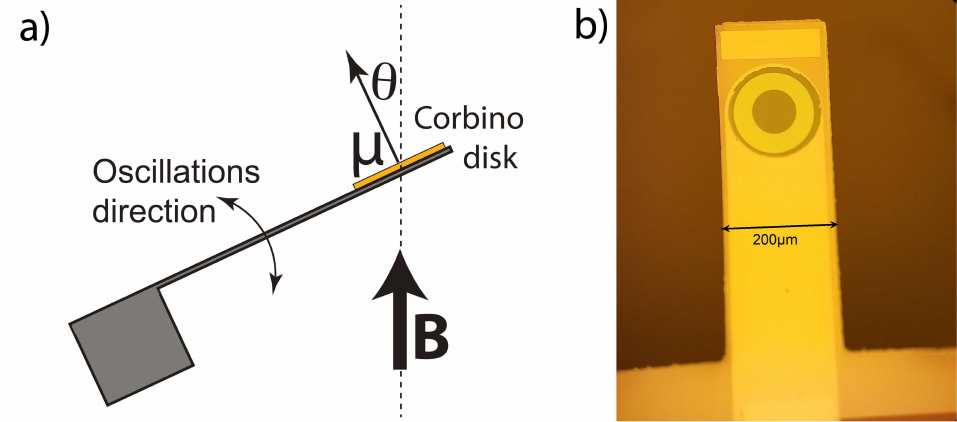}
        \caption{\label{CP} (a) Schematic of torque magnetometry where Hall currents in a Corbino disk induce the magnetic moment $\mu$. (b) Picture of the ITO Corbino disk cantilever. The center contact is connected to an underlying ground plane. The metal pad on the end of the cantilever is used for optical alignment.}
\end{figure}

Cantilever torque magnetometry utilizes a high-$Q$ resonator to detect the interaction between a magnetic dipole $\vec{\mu}$ and an external magnetic field $\vec{B}$ ~\cite{me,Resonant-TM,PERFETTI2017171, PhysRevB.64.014516, Bleszynski-Jayich272}, where the resulting torque is \cite{PERFETTI2017171}
\begin{equation}
 \vec{\tau} = \vec{\mu} \times \vec{B}.
 \end{equation}
A schematic depiction of cantilever torque magnetometry is shown in Fig.~\ref{CP}(a). The angular response $\theta$ of a cantilever with moment of inertia $A$, resonant frequency $\omega_{0}$, and quality factor $Q$ subject to an external torque $\tau$ may be approximated as a damped harmonic oscillator following~\cite{s8053497}
\begin{equation}\label{baseEQ}
    A\ddot{\theta} + \frac{A}{Q} \omega_{0}\dot{\theta} + A\omega_{0}^2 \theta = \tau.
\end{equation}
The torque of a dipole parallel to the cantilever surface norm and a static $z$-aligned magnetic field $B$ shifts resonant frequency as
\begin{equation}\label{fu}
    A\omega_{0}^2 \rightarrow A\omega_{0}^2 - \mu B\textrm{,\ \ \ \ \ \ or \ \ \ \ \ \ } \frac{\Delta \omega_{0}}{\omega_{0}} = \frac{\mu B}{2A\omega_{0}^2}.
\end{equation}
The circulating current in a Corbino disk \cite{Corbino} patterned on a cantilever forms such a magnetic dipole parallel to the cantilever surface norm \cite{me}. Finally, cantilever resonant frequency is observed interferometrically and driven through laser radiation pressure \cite{me}.

Cantilevers with ITO Corbino disks were examined for magnetization and magnetic transport effects. Silicon cantilevers of dimension $200\times600\times3~\mu$m were fabricated with Corbino disks as shown in Fig.~\ref{CP}(b). The resulting devices exhibited $A = 3\times 10^{-18}$~kg-m$^{2}$ and $f_{0} = \omega_{0}/2\pi = 14.616$~kHz. Magnetization is measured through the voltage-independent
\begin{equation}\label{magDel}
    \Delta f_{0}(B) = f_{0}(B) - f_{0}(B=0)
\end{equation}
and sample transport properties for Corbino disk voltage $V$ are measured through
\begin{equation}\label{delf}
    \delta f_{0}(\pm V, B) = f_{0}(+V, B) - f_{0}(-V, B).
\end{equation}
The Hall effect is measured through the even in $B$ component of $\delta f_{0}(\pm V, B)$ while the odd component of $\delta f_{0}(\pm V, B)$ is proportional to the misalignment between the inner and outer Corbino disk contacts and longitudinal current. As shown by Mumford {\it et al.} \cite{me}, using Eqn. \ref{fu} the even component of $\delta f_0(\pm V, B)$ yields Hall conductivity
\begin{equation}
\sigma_{xy}= C\frac{\delta f_{0}}{BV}\textrm{, where } C = \frac{4A f_0{\rm ln}(r_o/r_i)}{\pi (r_o^2-r_i^2)} 
\label{sigma}
\end{equation}
and $r_o$ and $r_i$ are the outer and inner radii of the disk. Thus, $\delta f_0(\pm V, B)$ is proportional to $\sigma_{xy}B$. 
\begin{figure}[h]
     \centering
         \includegraphics[width=0.4\textwidth]{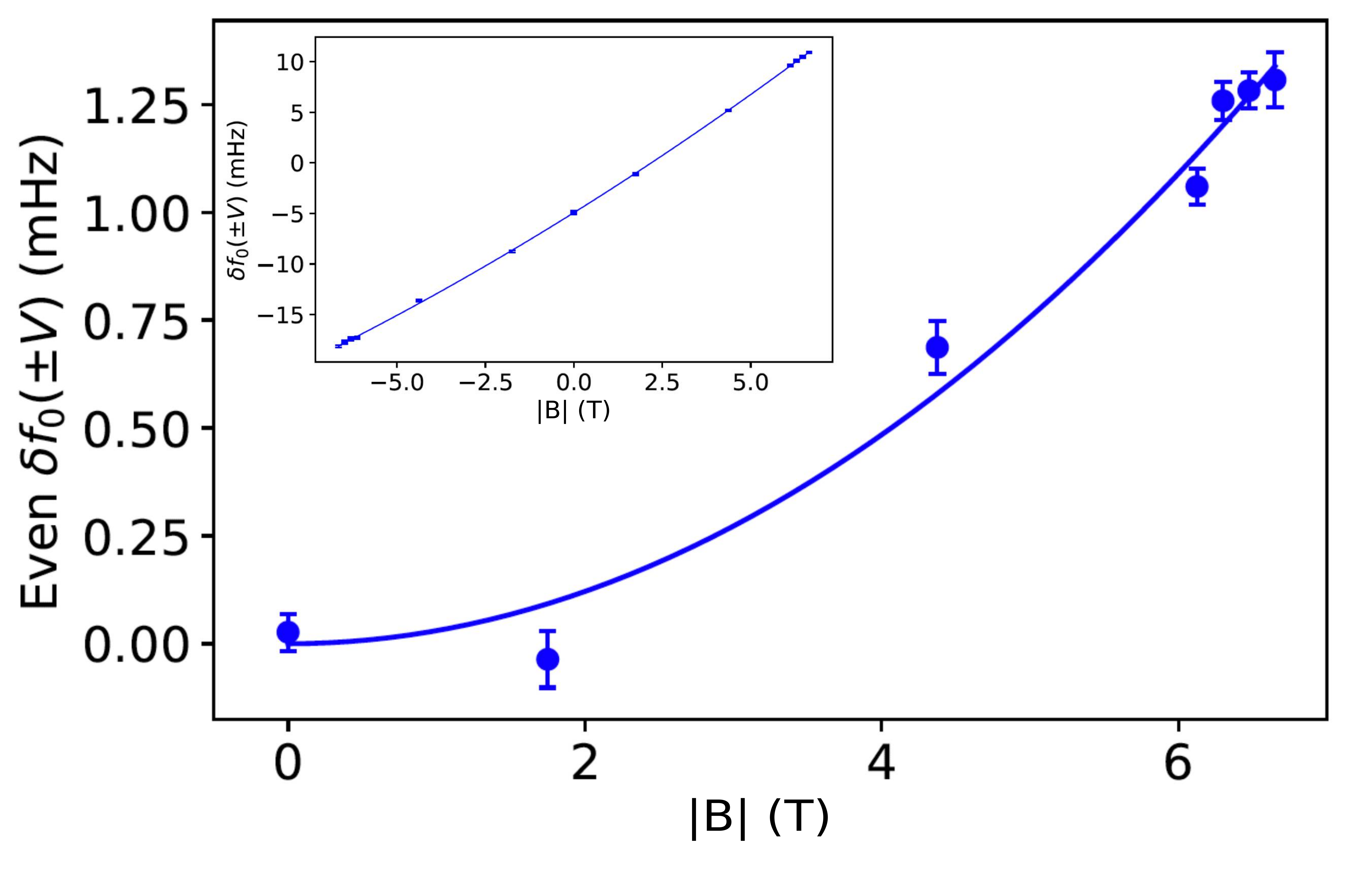}
        \caption{\label{EH} Example of extracting the even component of the frequency shift from which the Hall conductivity is calculated for non-magnetic metallic ITO. Inset shows the total frequency shift as a function of magnetic field.}
\end{figure}
The expected Hall contribution to $\delta f_0$ is proportional to $B^2$ as $\sigma_{xy}\propto B$. A typical $B^{2}$-like shift in $\delta f_0(\pm V, B)$ and background linear in $B$ shift in $\delta f_0(\pm V, B)$ due to misalignment are shown in Fig.~\ref{EH}.

\section{Results and Discussion}

\subsection{Characterization of Samples}
\begin{figure}[h]
     \centering
         \includegraphics[width=0.42\textwidth]{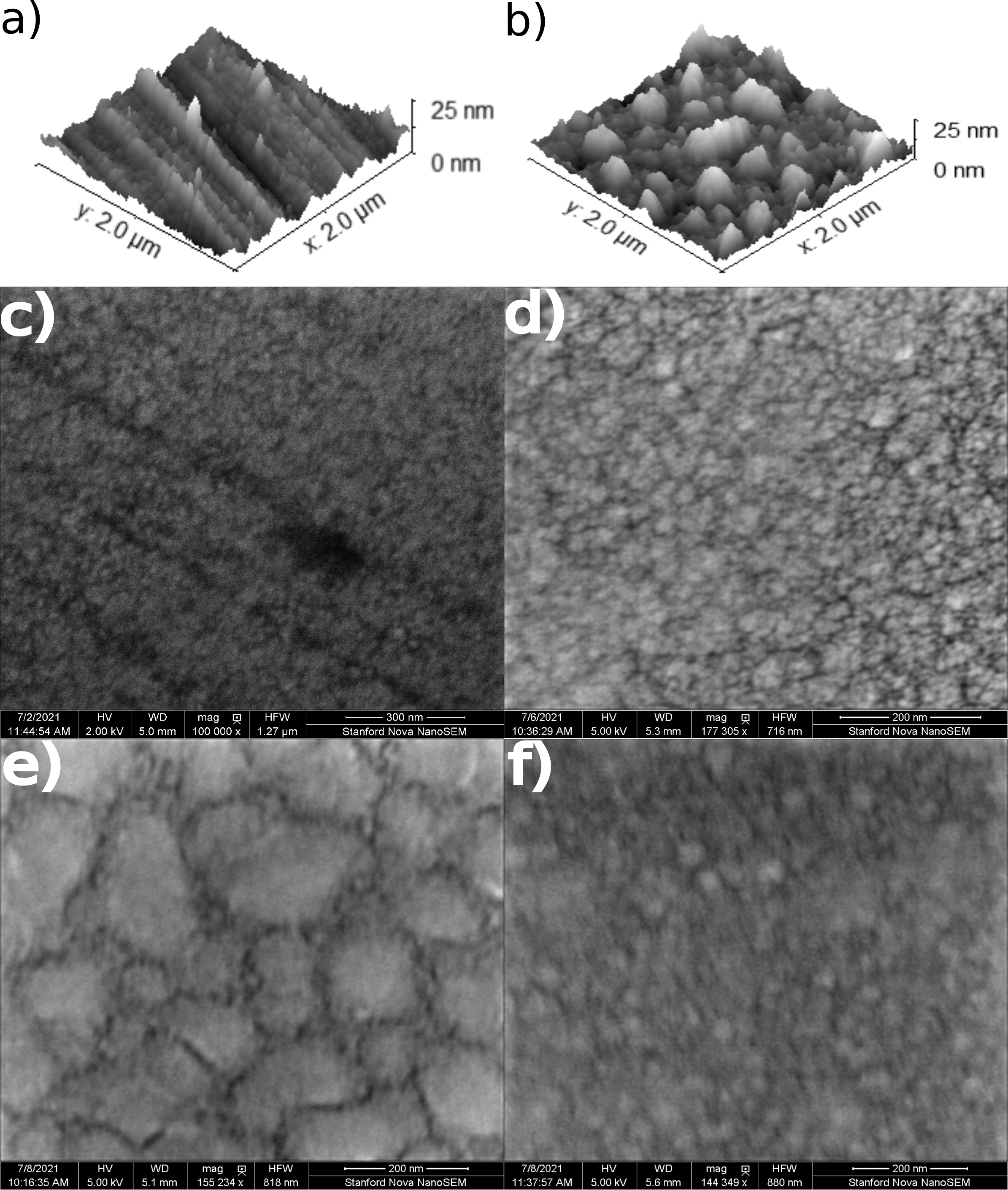}
        \caption{\label{grains} Characterization of the granularity of ITO with annealing. (a-b) AFM profile of ITO Hall bar before and after annealing. Sample height fluctuations and granularity increase after annealing. (c-d) SEM images of ITO Hall bar before and after annealing. (e-f) SEM images of Corbino disk ITO before and after annealing. The initially large patches of ITO or smooth ITO breaks down into granular areas after annealing.}
\end{figure}

\begin{figure}[h]
     \centering
         \includegraphics[width=.42\textwidth]{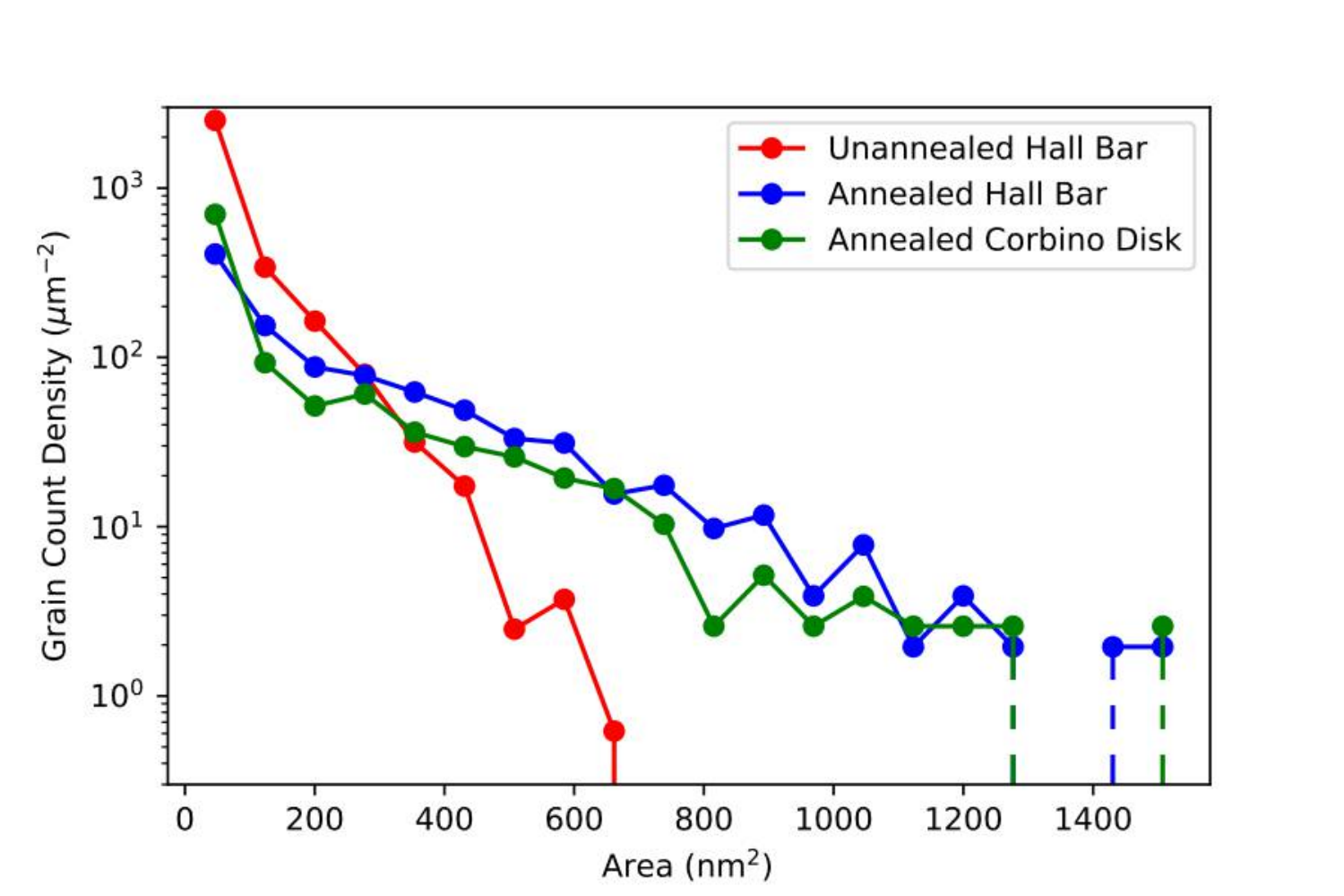}
        \caption{\label{GrainComp} Comparison of the grain sizes between the annealed and non-annealed ITO samples by SEM. Grains were identified by adaptive thresholding of the SEM images shown in Fig.~\ref{grains}. Fluctuations in SEM brightness changed from local noise to distinct larger grains with annealing.}
\end{figure}

ITO grain size and granularity increased with annealing as shown in Fig. \ref{grains}. As deposited, the ITO was largely amorphous with a smooth surface profiled by atomic force microscopy (AFM). After annealing, the ITO subdivided into grains with height fluctuations nearly equal to the deposited ITO thickness as seen in Fig. \ref{grains}(a-b). Complimentary scanning electron microscope (SEM) images also demonstrate a change in granularity with annealing. Two samples of ITO were analyzed by SEM. Sample 1 was patterned in a Hall bar and sample 2 was patterned in a Corbino disk. As deposited, the ITO exhibited either very small ITO grains or larger smooth ITO domains. The differences between the as-deposited ITO in the Corbino disk and Hall bar may be explained by a difference in deposited ITO morphology based on substrate \cite{substrate}. After annealing, the ITO became granular with typical grain size on the order of 500~nm$^{2}$. A clear increase in grain size with annealing is seen in Fig. \ref{GrainComp} which compares the grain distribution of unannealed and annealed ITO SEM images. The sheet resistance at base temperature also decreased with annealing, while the carrier density increased as is shown in Table~\ref{tab:my_label}.
Such a change in grain size \cite{SEM_paper, hakimi_ch4, RTA_ITO, Lee_grain,kerkache2007,AFM,gulen_2012}, decreased resistivity, and increased carrier density \cite{SEM_paper, wu_chiou_1993, RTA_ITO, Lee_grain,kerkache2007} are consistent with previous observations on ITO with higher temperature thermal annealing.


\begin{table}[h]
    \centering
    \begin{tabular}{|c|c|c|}
    \hline
         Resistivity ($k\Omega/\square$) & Configuration & $n$ (10$^{18}$ cm$^{-3}$) \\
         \hline\hline
         3300 & Corbino Disk & -0.08 $\pm$ 0.2 \\
         \hline
         460 & Corbino Disk & 0.17 $\pm$ 0.04 \\
         \hline
         185 & Corbino Disk & 1.7 $\pm$ 1\\
         \hline
         48 & Corbino Disk & 3.4 $\pm$ 0.5 \\
         \hline
         32 & Corbino Disk & -2.7 $\pm$ 0.3 \\
         \hline
         17 & Corbino Disk & 6.5 $\pm$ 1.4 \\
         \hline
         15 & Hall Bar & 9.33 $\pm$ 0.02 \\
         \hline
         8.5 & Corbino Disk & 22 $\pm$ 10 \\
         \hline
         4.6 & Hall Bar & 23.7 $\pm$ 0.1\\
         \hline
    \end{tabular}
      \caption{Samples investigated in this study and their transport properties at base temperature.}
    \label{tab:my_label}
\end{table}

\subsection{Corbino Disk Torque Magnetometry of ITO}

\subsubsection{Hall Conductivity Annealing Across MIT}

\begin{figure}
     \centering
         \includegraphics[width=.45\textwidth]{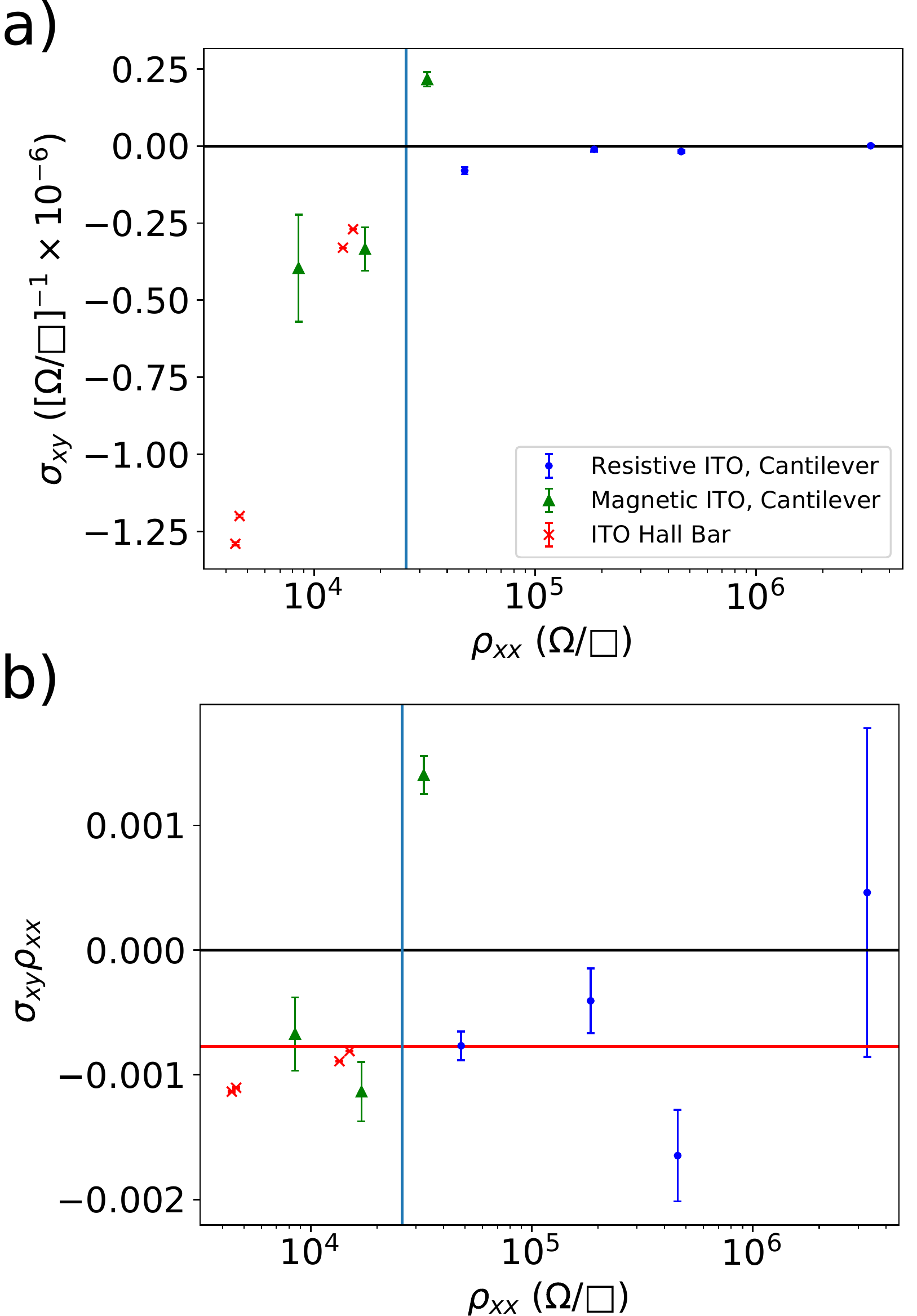}
        \caption{\label{bd} Behavior of $\sigma_{xy}\rho_{xx}$ at 5~T measured on a series of annealed ITO samples through the MIT. The outlying sample very close to the MIT may be strongly influenced by the magnetic behavior as discussed in Secs. \ref{MagMIT}-\ref{HC}. Additional data obtained through a Hall bar approach and discussed in Sec. \ref{HRT} is included as a reference.}
\end{figure}

An initially highly resistive ITO cantilever (resistivity of 3300 k$\Omega/\square$ at base temperature) was annealed in 6 steps to investigate the variation of Hall conductivity through the MIT as well as the onset and evolution of magnetism. Fig.~\ref{bd}(a) shows the extracted Hall conductivity of an ITO sample with annealing, while the respective carrier densities are given in Table~\ref{tab:my_label}. Annealed Hall-bar configuration samples are included for comparison. Ferromagnetism emerged close to the MIT and thus complicated extracting the ordinary Hall effect from $\delta f_{0}(\pm V, B)$ as discussed in further detail below. This increased complexity increases the uncertainty in $\sigma_{xy}$ extracted near the MIT, but the observed carrier densities remain consistent with similarly prepared samples measured in a Hall bar configuration.

An important result to be drawn from the high resistance samples deep in the hopping regime is the tendency of $\sigma_{xy}$ to vanish with increasing disorder or increasing $\rho_{xx}$. The hopping regime Hall effect is governed by the self-interference effect of the electron wave function propagating along limited paths dictated by magnetic field and the impurity sites responsible for the impurity band. Early theoretical work by Friedman and Pollak \cite{Pollak1981} using the Holstein approach \cite{Holstein1959} concluded that the Hall resistivity diverges as the temperature tends to zero. This result was then reinforced by Entin-Wohlman, {\it et al.} \cite{Imry1995}, who attempted to explain the discrepancy with complementary calculations predicting an electronic state where the longitudinal resistance diverges at zero temperature, but the Hall resistivity remains constant \cite{Viehweger1991,Kivelson1992,Imry1993}. Such a state is known as a Hall insulator. Indeed, while experimental results focused on the vicinity of the MIT confirmed divergence of $\rho_{xy}$ on the insulating side \cite{Hopkins1989,Koon1990}, in other experiments insulating states derived from quantum Hall states were consistent with a Hall insulator. The result for $\rho_{xy}$ may be different depending on the order of temperature and frequency limits taken in the DC limit of the Hall resistivity. Similarly, it was emphasized in \cite{Imry1995} that if both $\rho_{xy}$ and $\rho_{xx}$ are calculated and averaged over the disorder, they should both show divergence, while if the respective conductivities are averaged over the disorder the calculated $\rho_{xy}$ may approach a constant. We directly measure $\sigma_{xy}$ and $\rho_{xx}$, which is a hybrid of the above two procedures.

We show $\sigma_{xy}\rho_{xx}$ in Fig.~\ref{EH}(b) to test the Hall behavior. The observed $\sigma_{xy}\rho_{xx}$ is approximately a constant, excluding the very vicinity of the MIT. Far above the MIT weak localization corrections may alter the ratio \cite{Koon1990} at a level smaller than the resolution of the experiment. However, on the insulating side, a Hall insulating behavior would require that $\sigma_{xy}\rho_{xx}\sim\rho_{xx}^{-1}$ in contrast to the observed behavior. Such behavior supports that in the hoping regime $\rho_{xy}\propto\rho_{xx}$, or that both will show similar divergence as $T\to 0$. 

\subsubsection{Magnetism Near MIT}\label{MagMIT}

Successive annealing of highly insulating ITO resulted in increased granularity, lower resistivity, and, as the MIT is approached, the emergence of ferromagnetism. We will now profile the origin of the surprising magnetism, as well as its anisotropy and dependence on the film morphology, carrier density, and defect density. 
\begin{figure}[h]
     \centering
         \includegraphics[width=0.4\textwidth]{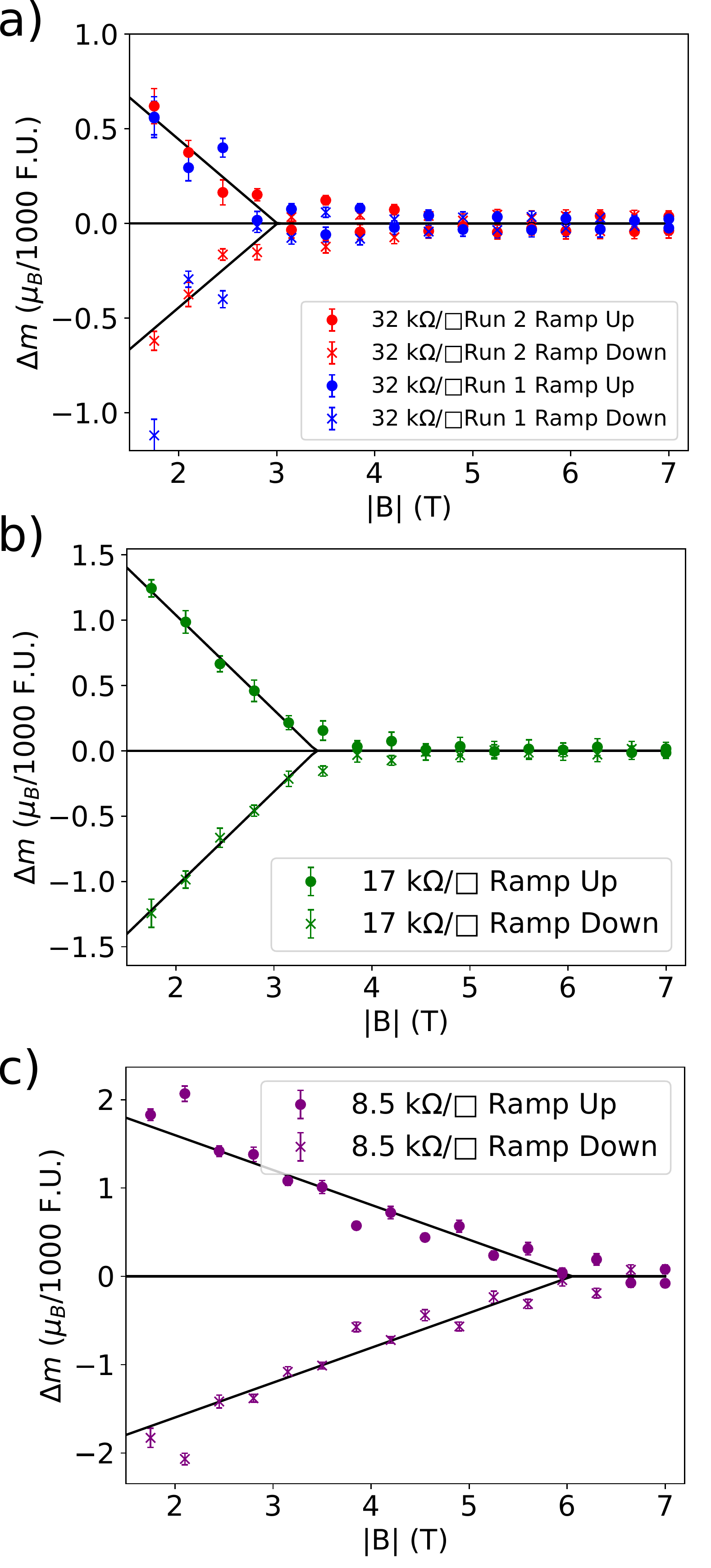}
        \caption{\label{Mags} Hysteresis in magnetization $\Delta m$ attributed to magnetization for three annealed ITO resistivities. We calculate the hysteretic magnetic moment per ``formula unit'' (f.u.), approximated as the unit cell volume of crystalline In$_2$O$_3$. The amplitude of hysteresis and saturating field increase with with longer annealing as shown in Table \ref{MagTab}. Ramping in magnetic field to $\pm~7~$T was performed in 0.7~T steps.}
\end{figure}

Anisotropy in bulk magnetization may be observed through cantilever torque magnetometry \cite{Resonant-TM, PERFETTI2017171, PhysRevB.64.014516, Bleszynski-Jayich272}. Hysteresis in magnetization due to ferromagnetic ordering results in hysteresis in $\Delta f_{0}(B)$ observed through Eqns. \ref{fu} and \ref{magDel}. Hysteresis in magnetization arose from ferromagnetism after annealing near the MIT transition point as shown in Fig. \ref{Mags}. All tested cantilevers with patterned metallic layers displayed low-field shifts in $f_{0}$ as discussed in Appendix \ref{LowB}, thus only the hysteresis in magnetization for $|B| > 1.5~$T can be attributed to the ITO. The amplitude of the magnetic hysteresis loop increased with annealing and confirms that the observed $\Delta f_{0}$ hysteresis arose from changes in the ITO. To rule out systematic factors, the order of data-taking was switched between the two 32~k$\Omega/\square$ datasets, and a 6 hour delay at $+7~$T was inserted between the sweep up and sweep down runs for the 17~k$\Omega/\square$ dataset. The lack of hysteresis at high magnetic field confirms that hysteresis in $\Delta f_{0}(B)$ arose from changes in the ITO instead of systematic factors. No hysteresis in magnetization was found for ITO with sheet  resistance $> 32~$k$\Omega/\square$.

\begin{table}
    \centering
    \begin{tabular}{|c|c|c|}
    \hline
         Resistivity (k$\Omega/\square$) & $B_{S}$ (T) & $\alpha$ (mHz/T$^2$) \\
         \hline
         32 & 3$\pm$0.1 & 3.6$\pm$0.7\\
         \hline
         17 & 3.6$\pm$0.1 & 5.4$\pm$0.5 \\
         \hline
         8.5 & 6$\pm$0.2 & 3.1$\pm$0.3 \\
         \hline
    \end{tabular}
    \caption{\label{MagTab} Hysteretic magnetization fit coefficients obtained through torque magnetometry. Magnetization was more pronounced as $B_{s}$ increased with increased annealing and lower resistance}
\end{table}

Magnetization hysteresis or hysteresis in $\Delta f_{0}$ above 1.5~T was fit to $\Delta f_{0, hyst} = \alpha(B - B_{S})B$ for saturating field $B_{S}$ and proportionality constant $\alpha$. Fit coefficients are provided in Table \ref{MagTab}. The amplitude of magnetization and $B_{S}$ increased with annealing, as previously observed in ITO \cite{hakimi_paper}. Note that stronger $z$-axis ferromagnetism raises $f_{0}$, while in-plane magnetism can decrease $f_{0}$. The lower $f_{0}$ during the ramp down in magnetic field is thus consistent with in-plane ferromagnetism. In-plane ferromagnetism has been observed previously in ITO, and the higher training fields are consistent with a largely out-of-plane applied field instead of a purely in-plane field \cite{hakimi_paper, hakimi_ch5}.

The most annealed 8.5 k$\Omega/\square$ sample displayed a hysteretic magnetization saturation of $\Delta m\approx 0.002 \mu_B/{\rm f.u.}$ where f.u. is the In$_2$O$_3$ unit cell volume. If we assume that the associated moment with each defect is of order $\sim1\mu_B$, we obtain an approximate magnetic defect density of $1.3\times 10^{19}$ defects/cm$^3$, within a factor of 2 of the carrier density shown in Table \ref{tab:my_label}. However, the strong in-plane anisotropy observed in AHE measurements as well as the large anisotropy field suggest that the relevant oxygen defects and vacancies concentrate at surfaces and interfaces, presumably due to the granular structure as shown in Fig.~\ref{grains}. If we distribute the moments only at the surface of the film, and assume that because of surface roughness the effective surface area is increased by a factor of two, we obtain a surface magnetization of $\sim 0.1~\mu_B/$f.u.. One out of 10 unit cells at the surface introduces a magnetic moment if each surface defect produces $\sim1\mu_B$. Finally, the anisotropy energy $K_u$ can be approximated from the saturation magnetization and field. The approximate $K_u =B_s\Delta m_s/2 \approx 2\times 10^4 {\rm ~J/m}^3$ is about 10 times smaller than a typical anisotropy energy of a thin film of polycrystalline ferromagnetic Fe$_3$O$_4$ of similar thickness \cite{Liu_2015}.

\subsubsection{Anomalous Hall Conductivity Near MIT}\label{HC}

The most direct demonstration of intrinsic ferromagnetism is the emergence of the anomalous Hall effect in the transverse channel.  Using Corbino disk torque magnetometry, we can detect both the Hall and anomalous Hall effects. However, unlike a standard Hall bar configuration, the sample is voltage-biased and transverse conductivity $\sigma_{xy}(B)$ is directly measured by the magnetic moment created by circulating Hall currents.
\begin{figure}
     \centering
         \includegraphics[width=0.4\textwidth]{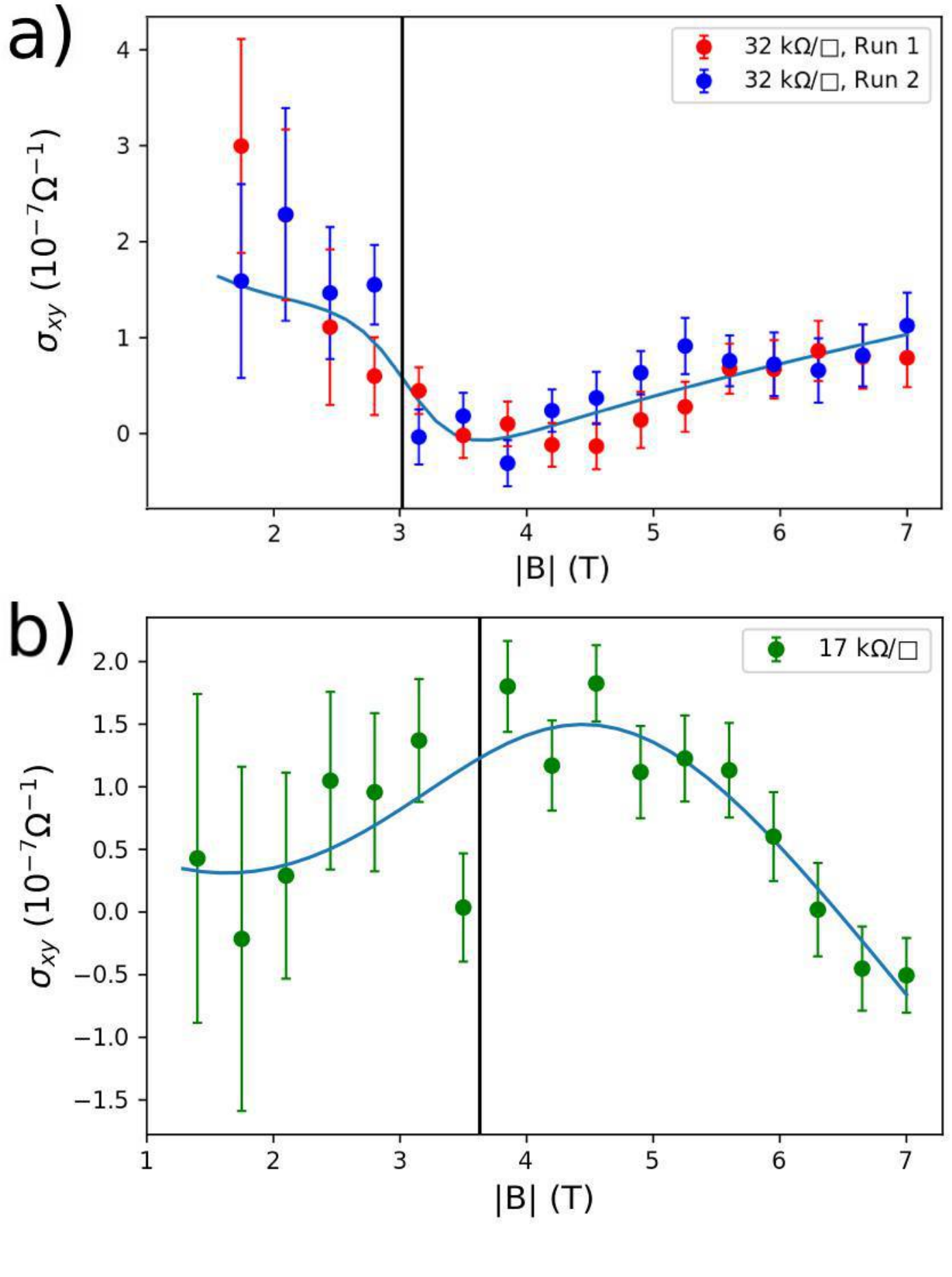}
        \caption{\label{EVs} Fits to the even component of $\delta f_{0}(\pm V, B)$ assuming a sigmoidal kink in anomalous Hall signal at $B_{s}$. For comparison, $B_{s}$ as fit from the closing of the $\Delta f_{0}$ hysteresis in Fig. \ref{Mags} is shown with a vertical line.}
\end{figure}
For a material with $z$-component of magnetization $M_{z}$ and coefficients $a$ and $b$, 
\begin{equation}\label{meq}
    \sigma_{xy} = aB + bM_{z} \rightarrow \delta f_{0}(\pm V, B) \propto B^{2} + \beta BM_{z}.
\end{equation}
The extracted and fit $\sigma_{xy}$ from $\delta f_{0}(\pm V, B)$ is shown in Fig. \ref{EVs}. The even $\delta f_{0}(\pm V, B)$ is fit to Eqn. \ref{meq} assuming a diamagnetic sigmoidal $M_{z}$ appearing at the saturating field listed in Table \ref{MagTab}. Such behavior would be expected from a magnetic transition from an in-plane ferromagnet to an out of plane paramagnet. The region of varying anomalous Hall signal corresponds with the hysteresis loop bounds shown in Fig. \ref{Mags} as expected. The amplitude of the anomalous Hall signal also increased in the more conductive device as is expected based on the increased strength of magnetism with annealing.

Finally we comment on the sign change of the 32 k$\Omega/\square$ sample Hall conductivity shown in Fig.~\ref{EVs}(a). Here we note that this sample was close to but on the insulating side of the MIT and thus at the onset the variable range hopping regime. A sign change of the anomalous Hall effect is then possible as a function of the impurity band filling \cite{Burkov_2003}. As our films experienced strong variations in oxygen vacancies and defects as the morphology changed through annealing, such an effect may be anticipated. Indeed, such sign reversal was previously observed in (In$_{0.27}$Co$_{0.73}$)$_2$O$_{3-v}$ ($v$ denotes the oxygen vacancies) ferromagnetic semiconductors \cite{qiao_2014} in the VRH regime, particularly as the temperature was reduced and localization effects dominate. As materials like ITO are based on the In$_2$O$_3$ oxide system where oxygen vacancies and morphology contribute to the creation of the impurity band, the observation of sign change in our ITO samples may have the same origin.

\subsubsection{Hysteretic Transport}
\begin{figure}
     \centering
         \includegraphics[width=0.4\textwidth]{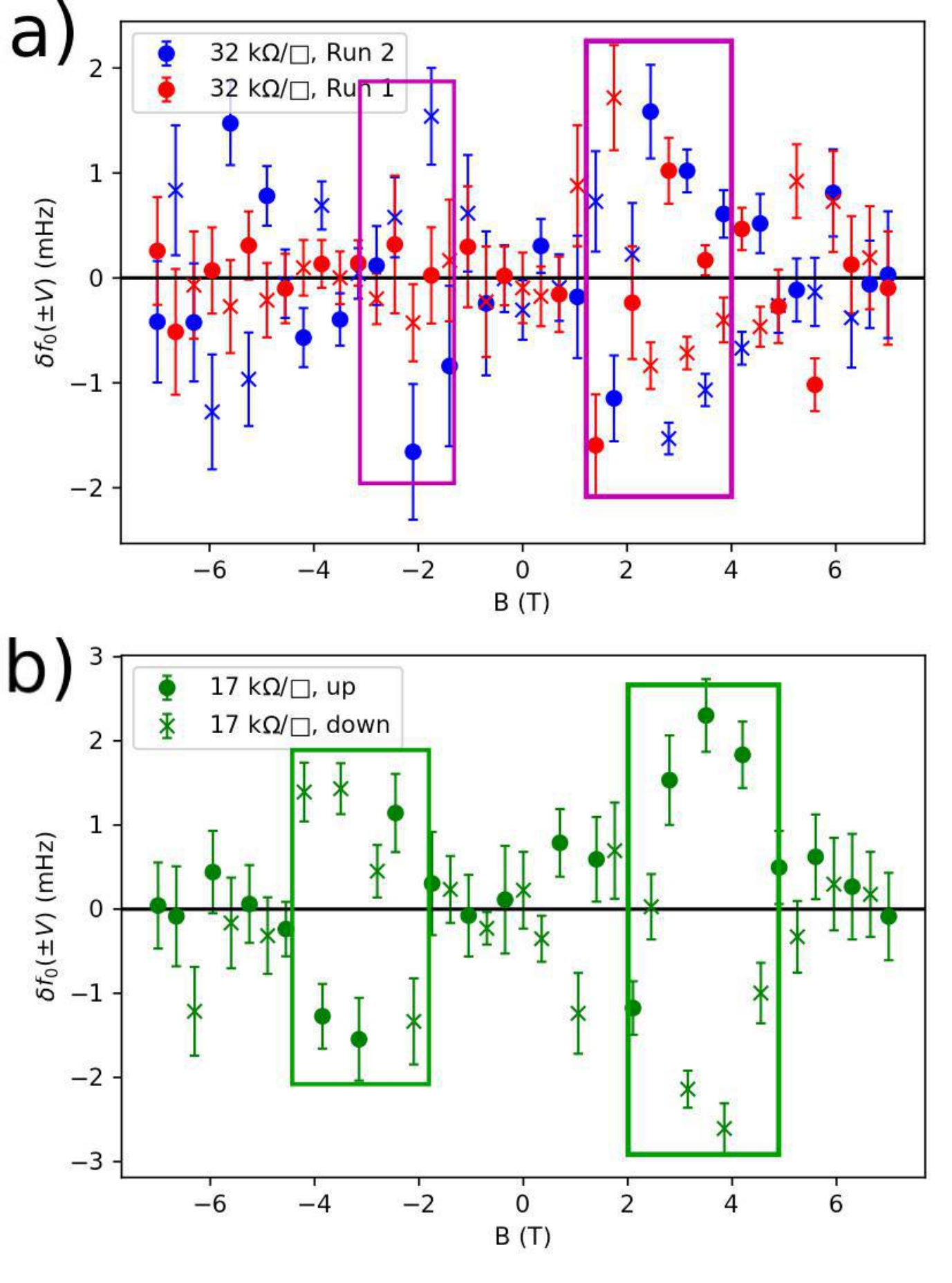}
        \caption{\label{HistTrans} Hysteresis in transport or $\delta f_{0}(\pm V, B)$ for 32~k$\Omega/\square$ and 17~k$\Omega/\square$ annealed devices. Peaks near the closing of the hysteresis loops are highlighted in boxes. The Corbino disk voltage was $\pm 0.3$~V. Hysteresis could appear due to both magnetoresistance for $B \lesssim B_{s}$ and the anomalous Hall effect as $B \approx B_{s}$.}
\end{figure}
The detection of AHE is also accompanied by hysteresis in $\delta f_{0}(\pm V, B)$. As shown in Fig. \ref{HistTrans}, there is a hysteretic component of $\delta f_{0}(\pm V, B)$ with peaks coinciding with the closing of the $f_{0}$ hysteresis loops. The smoothly varying non-hysteretic component of $\delta f_{0}(\pm V, B)$ is the average between the up and down magnetic field sweep values of $\delta f_{0}(\pm V, B)$ and is used to extract Hall conductivity and magnetoresistance. The hysteretic component of $\delta f_{0}(\pm V, B)$ is the difference between the smoothly varying background and $\delta f_{0}(\pm V, B)$ in each sweep direction. Hysteresis in transport closes above $B_{S}$ and qualitatively peaks near $B_{S}$ for annealed magnetic ITO cantilever datasets. The hysteretic $\delta f_{0}(\pm V, B)$ is approximately 0 at low field, although such a closing may be due to the lack of dipole sensitivity at $B=0$. While hysteretic $\delta f_{0}(\pm V, B)$ signal cannot be simply ascribed to a single source as both magnetoresistance and Hall conductivity have contributions from sample magnetization, the fact that the hysteresis profile is not entirely even or odd in $B$ suggests that both components are hysteretic. Finally, the $\delta f_{0}(\pm V, B)$ hysteresis coincided with the hysteresis in magnetization and its evolution with annealing, confirming that both the transport and magnetization hysteresis originate from the ITO.

\subsection{ITO Hall Bar Transport}\label{HRT}

To confirm our findings using Corbino disk torque magnetometry and highlight the power of the technique, we performed complementary measurements using a standard Hall bar configuration. For a given temperature, the sample was current-biased, and the longitudinal and transverse voltages were recorded as a function of magnetic field.  However, detection of the anomalous Hall effect in the transverse channel in such a configuration may be difficult due to sample issues such as strong in-plane anisotropy or high sample resistance \cite{kim_ji_2006,hakimi_ch5,Stankiewicz_2007}. Negative magnetoresistance due to the reduction in spin-related scattering thus is typically used to relate transport to direct magnetization measurements in ITO. Hysteretic magnetoresistance would provide convincing evidence of ferromagnetism, pointing to the presence of magnetic domains that need to be flipped in direction when the magnetic field is reversed.
\begin{figure}
     \centering
         \includegraphics[width=.45\textwidth]{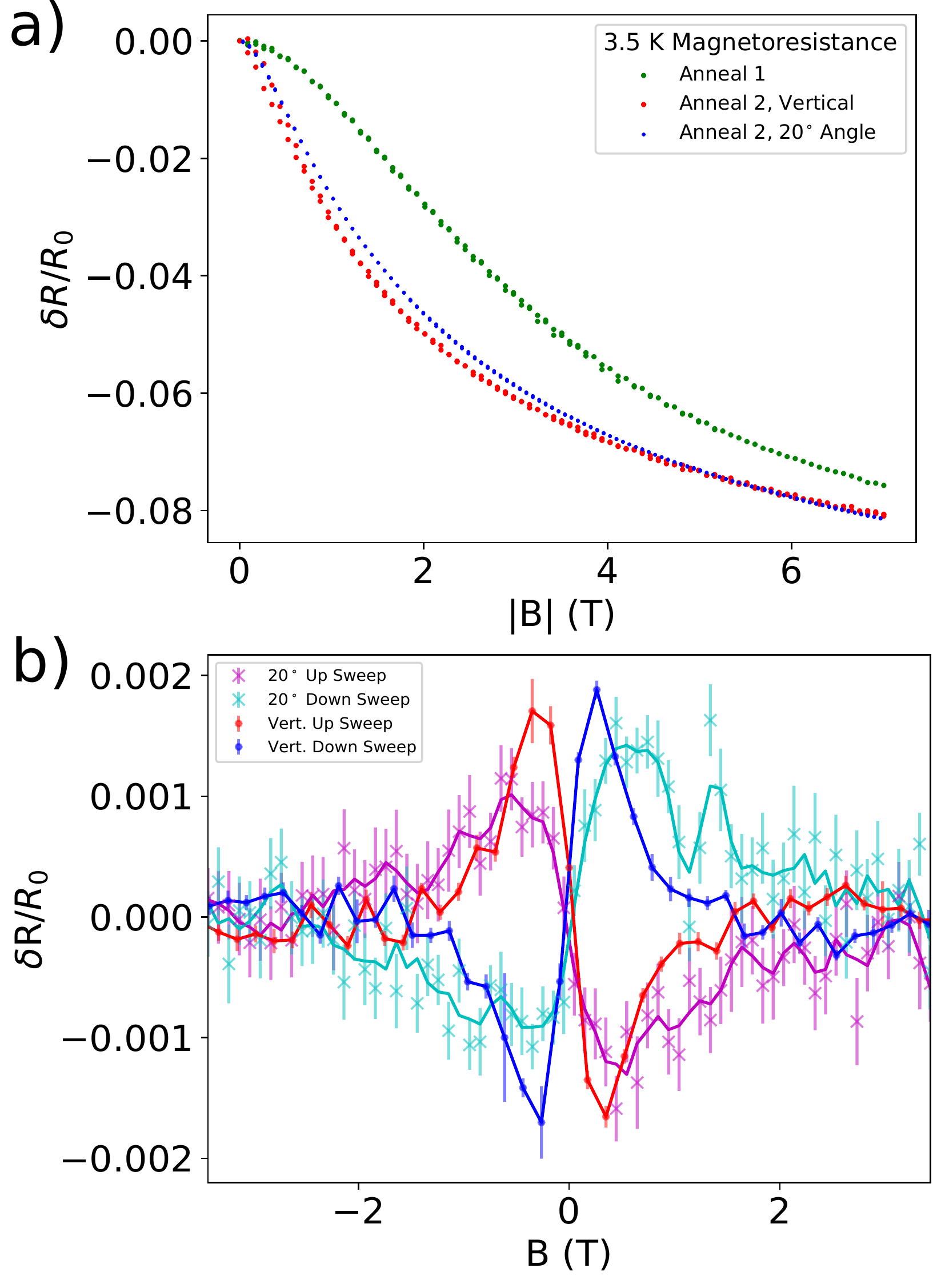}
        \caption{\label{MRHB} (a) Magnetoresistance for a lightly annealed and more strongly annealed ITO sample. (b) Hysteresis in magnetoresistance after two stages of annealing for a vertically and 20$^{\circ}$ mounted Hall bar. Hysteresis is slightly more pronounced for the vertically mounted sample. Magnetoresistance was observed by a continuous sweep in $B$ for the 20$^{\circ}$ angle mounted dataset and by stopping at set $B$ for the vertically mounted dataset.}
\end{figure}

Initially resistive ITO was sputtered in a Hall bar pattern and successively annealed to detect the onset of magnetism. Note that Hall bar measurements were performed on more conductive ITO films of the same thickness as the Corbino disk samples due to the difficulties in measuring high resistance samples using a Hall bar technique. Relative magnetoresistance or $\delta R/R(0)\equiv [R(B)-R(0)]/R(0)$ is shown in Fig. \ref{MRHB}(a). The $B=0$ resistance was 15~k$\Omega/\square$ for the first anneal and 4.6~k$\Omega/\square$ for the second anneal. More strongly annealed or conductive samples displayed more dramatic decreases in $\delta R/R_{0}$ at $|B| < 2~$T and asymptotic $\delta R/R_{0}$ for $|B| > 5~$T. Such changes with annealing are consistent with previous studies of ITO with changing carrier density and temperature \cite{fujimoto}. Magnetoresistance of the ITO Hall bar mounted at a 20$^{\circ}$ angle is also shown in Fig.~\ref{MRHB}(a) as a test for anisotropic magnetoresistance. However, any changes in $\delta R/R_{0}$ with mounting angle were minimal and could be explained by slight differences in annealing between the vertically and angle-mounted Hall bars. Resistivity and carrier density of the Hall bar samples are included in Table~\ref{tab:my_label}.

Hall effect and magnetoresistance data were further examined for evidence magnetism. Hysteretic magnetoresistance below 2~T for both the vertically mounted and angled ITO Hall bars after annealing is shown in Fig.~\ref{MRHB}(b). The sign of hysteresis is consistent with previously observed in-plane magnetic ordering \cite{majmuder, hakimi_paper, hong_2005} and appeared only with annealing or increased granularity.  No hysteretic magnetoresistance was observed in the more lightly annealed dataset likely due to either increased magnetism with annealing or a lower slope in $\delta R/R(0)$ near $B=0$. No discernible anomalous Hall contribution could be detected, and the absence of AHE is consistent with strong in-plane magnetic anisotropy.

The Corbino disk cantilevers exhibit significantly higher $B_{s}$ than the hysteresis-closing $B$ in the Hall bar samples shown in Fig.~\ref{MRHB}(b). The saturating field observed in the more conductive Hall bar ITO sample is lower than any of the saturating fields of the magnetic Corbino disk samples but is closest to that of the more lightly annealed 32~k$\Omega/\square$ samples. Magnetization in ITO thus is not dictated purely by longitudinal resistivity and results from an interplay between granularity and carrier density.

\section{Conclusions}

We have shown unambiguous evidence of in-plane ferromagnetism and its effect on the transport properties of low carrier density ITO annealed through its MIT. Using a novel Corbino disk torque magnetometry technique \cite{me}, hysteresis in the voltage-independent shift in $f_{0}$ with $B$ provides a direct measurement of in-plane magnetic ordering, while at a finite voltage $\delta f_{0}(\pm V, B)$ exhibits AHE and hysteretic transport. In particular, the observation of AHE provides direct evidence of inherent magnetism in ITO. Examination of both the Hall and anomalous Hall contributions to $\delta f_{0}(\pm V, B)$ confirm the breaking of an in-plane magnetic ordering. Magnetism arose with annealing in ITO and thus with changes to the magnetic oxide morphology. The observation of intrinsic magnetism in ITO correlated with changes in surface morphology supports the hypothesis that magnetism arises from surface oxygen vacancies. Such Hall and magnetization measurements may be performed for deeply insulating materials due to the direct measurement of $\sigma_{xy}$. Initial results of $\sigma_{xy}$ in resistive ITO contradict expectations for both a Hall insulator and standard models of Mott variable range hopping \cite{HCVRH, koon_castner_1987, gruenewald_mueller_thomas_wuertz_1981}. Additionally, Corbino disk torque magnetometry provides the ability to simultaneously measure the bulk magnetic and transport properties of a material across the MIT, opening avenues for new physics.

\begin{acknowledgments}
This work was funded by the Army Research Office grant W911NF1710588 and by the Department of Energy, Office of Science, Basic Energy Sciences, Materials Sciences and Engineering Division, under Contract DEAC02-76SF00515.
\end{acknowledgments}

\appendix

\section{Low Field $f_{0}$ Dependence}\label{LowB}
\begin{figure}[h]
     \centering
         \includegraphics[width=.45\textwidth]{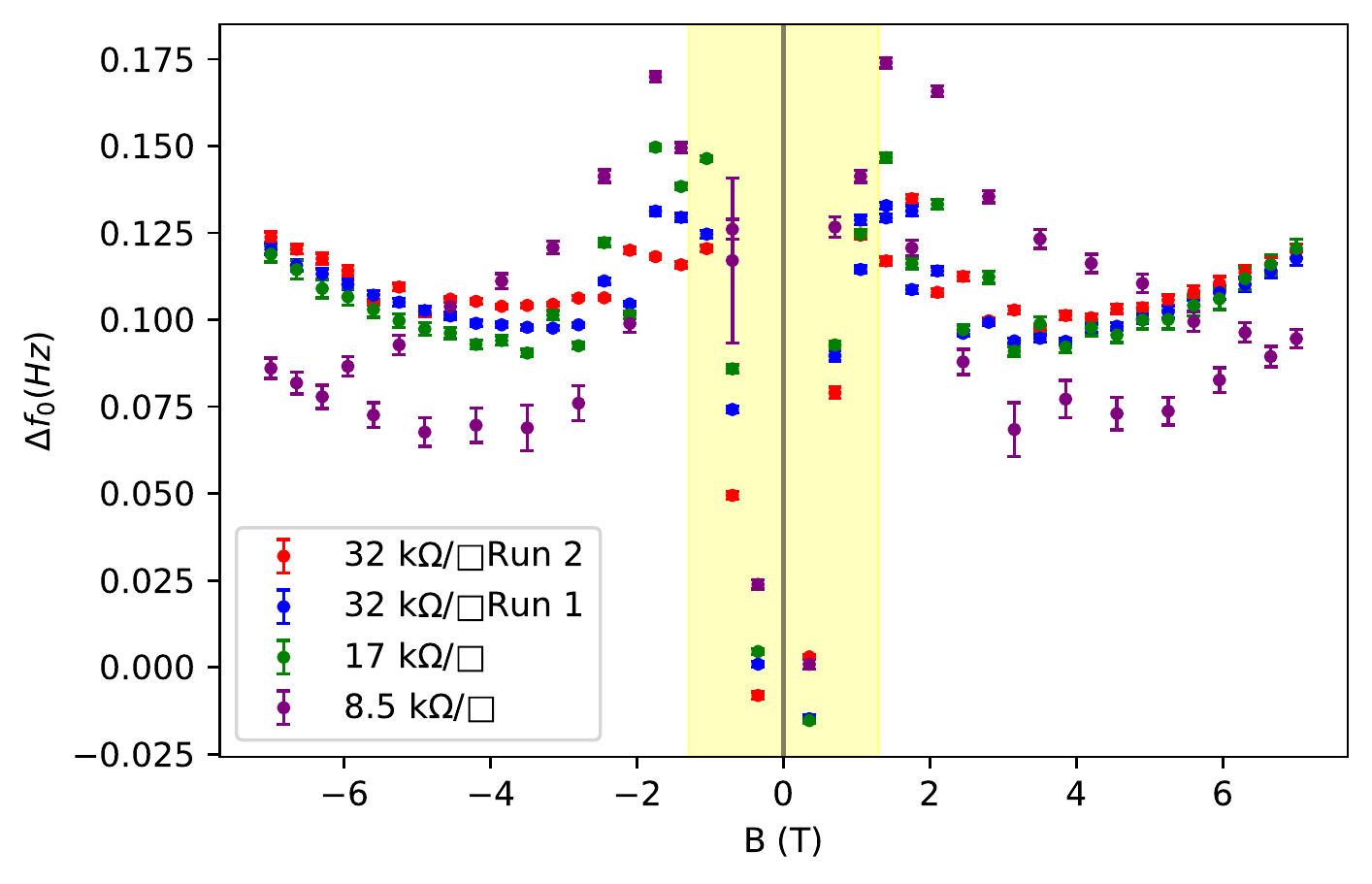}
        \caption{\label{f0s} Voltage-independent $\Delta f_{0}$ for four cooldowns. There is rapid increase in $\Delta f_{0}$ in the yellow highlighted region of $B$ near 0~T seen in all cooldowns regardless of sample magnetization. The low field $\Delta f_{0}$ is therefore not attributed to the ITO and is not fit in Fig. \ref{Mags}.}
\end{figure}

The full profile of $\Delta f_{0}$ with magnetic field for 4 annealing strengths is shown in Fig. \ref{f0s}. Hysteresis appears as fluctuations in $\Delta f_{0}$ between neighboring $B$ points when both sweep directions are plotted together. Without subtracting the background change in $f_{0}$ such hysteresis is most clear in the 8.5~k$\Omega$ sample.

\bigskip
\bibliography{bibl.bib}
\bigskip

\end{document}